\documentclass[manuscript]{aastex}
\usepackage{amsmath}
\usepackage{amssymb}
\usepackage{txfonts}
\usepackage{pdflscape}
\usepackage[colorlinks,
            linkcolor=blue,
            anchorcolor=blue,
            citecolor=black
            ]{hyperref}

\shortauthors{Yang et al.}

\begin{document}

\title{Relative Alpha in the Magneto-Hydro-Dynamics (MHD) with open magnetic field boundary and its application to the solar eruption}
\author{Shangbin Yang \altaffilmark{1,2,3}, J\"org
B\"uchner \altaffilmark{2,4}, Jean Carlo Santos \altaffilmark{5}, Jan Sk\'ala\altaffilmark{6} AND Hongqi Zhang \altaffilmark{1}}

\altaffiltext{1}{Key Laboratory of Solar Activity, National
Astronomical Observatories, Chinese Academy of Sciences, 100101
Beijing, China}

\altaffiltext{2}{ Max-Planck Institute for Solar System Research, 37077 G\"ottingen, Germany}

\altaffiltext{3}{University of Chinese Academy of Sciences, 100049 Beijing, PR China}

\altaffiltext{4}{Center for Astronomy and Astrophysics, Berlin Institute of Technology, 10623 Berlin, Germany}

\altaffiltext{5}{Universidade Tecnol\'ogica Federal do Paran\'a, Curitiba, Paran\'a, Brazil}

\altaffiltext{6}{Astronomical Institute of Czech Academy of Sciences, 25165 Ondrejov, Czech Republic}

\begin{abstract}
{An instability criterion in the MHD with the open boundary of magnetic field is proposed in this paper. We use a series of linear force-free extrapolation field, in which the normal part of magnetic field is fixed, to obtain the linear fitting coefficient called relative alpha by using the co-joined value of magnetic free energy and magnetic flux at the open boundary  ($E_f \Phi ^2$) and the square of relative magnetic helicity ($H_R^2$). We calculate this coefficient of the magnetic field above active regions NOAA~8210 and NOAA~11429 obtained by the photospheric-data-driven magnetohydrodynamics (MHD) model. It is found that the fitting coefficient is a good proxy of the criterion to indicate the occurrence of instability after which the magnetic reconnection happens and caused the fast release of magnetic energy. We also applied this method to the continuous evolution of three-dimension magnetic
field of NOAA~11158 based on the measurement of photospheric vector magnetic field of SDO/HMI by the Non-linear Force-Free (NLFF) extrapolation method.  The calculated coefficient when the major flare happened based on the extrapolation data is very close to the expected ones, which perfectly reflects the occurrence of instability and the difference is even less than 7\%. This relative alpha  is very helpful to evaluate how far it is from the instability in the MHD and quantitatively estimate the occurrence of solar eruption in the space weather forecast.
}
\end{abstract}


\section{Introduction}
It is well known that solar flares and coronal mass ejections (CMEs) are related to the sudden release of stored magnetic energy \citep[see, e.g.][]{Priest2002}. How are the corresponding instabilities triggered has become the central question to understand the above phenomena.  At large scales, where the magnetic Reynolds number is large enough, mechanisms such as kink instability \citep{TK05} ,  torus instability \citep{KT06} or ballooning instability \citep[e.g.][]{Fon01} could play an important role. At small scales, the magnetic Reynolds number becomes small enough such that the complexity of wave-particle interaction could efficiently cause the instability in the turbulent plasma \citep{Bun58}. Due to the nonlinear characteristics of the instability, analytical treatments usually fail and numerical simulation also could not cover the long time and infinite spatial frequency and time frequency. Note that CMEs and solar flares are both transient phenomena and last for a short time compared to the lifetime of a solar active region and, out of the period of energy release, the plasma system could be considered at a quasi-equilibrium state. It would be of great help to understand how the magnetic energy release occurs in the solar corona if we could evaluate the quasi-equilibrium, which corresponds to a MHD extreme state.

\cite{Woltjer1958} proposed that magnetic helicity in a closed system is conserved in the ideal MHD case. He also proposed that force-free fields with constant $\alpha$ represent the lowest state of magnetic energy which a closed system may attain by using variation principle of magnetic energy plus magnetic helicity, with a Lagrangian multiplier to find the MHD extreme state. In a general way, the extrema of $H+H_m$, which serves as a Lyapunov functional (also called Casimir functional) in an infinite-dimensional generalization of Dirichlet’s principle of Hamiltonian mechanics, corresponds to equilibria \citep[see eg.][]{And2010}. \cite{Berger1984} firstly proposed the concept of relative magnetic helicity to solve the gauge problem of vector potential of the magnetic field with open boundary. \cite{Berger1985} analyzed the relative magnetic helicity and magnetic energy in the constant $\alpha$ fields.
It is found that there may exist several $\alpha$ eigenvalues and the lowest such $\alpha$ is stable to all helicity preserving fixed-boundary perturbations. For most higher values of $\alpha$, there does exist an infinitesimal helicity-preserving perturbation which decreases the magnetic energy. This early work revealed the importance of the stability of $\alpha$ force-free field based on the analysis of the magnetic helicity and the magnetic energy with open boundary conditions. For the specific topological structures, the relation between energy and helicity of magnetic torus knots and braids in the closed boundary has been well studied recently by \cite{ObRicca18}, which revealed that magnetic energy is linearly related to helicity and the magnetic energy is bounded by the helicity.  However, the relation between magnetic energy and relative magnetic helicity in the MHD extreme state with open magnetic field boundaries in the solar magnetic field observations and MHD simulations is still not clear.  \cite{Yang13,Yang18} developed a method for calculating the relative magnetic helicity in a finite 3-D volume. The method has already been compared with other method in the review of \citet{Val16} and in this work we will use this method to calculate the relative magnetic helicity in the given 3D magnetic field. In this article, we propose an instability criterion coefficient, called relative alpha, estimated using the magnetic free energy, magnetic flux and relative magnetic helicity. In Sec.~\ref{sec:EFHR} we describe how to calculate the relative alpha in the open magnetic field  boundary by using an analytical force-free model \citep{Low90}. In Sec.~\ref{sec:test} we investigate the relationship between the relative alpha and the solar eruption by using 3D magnetic field data obtained by a data-driven MHD model \citep[GOEMHD3,][]{Skala15} for the simulation of NOAA 8210 and NOAA 11429,  also using NLFF extrapolation data of NOAA 11158 from the SDO/HMI. In Sec.~\ref{sec:summary}, we present the summary and discussion.

\section{Definition of Relative Alpha}
\label{sec:EFHR}

The basic idea is that for a force-free magnetic field, satisfying $\nabla \times \bf B=\alpha \bf B$, the condition $\nabla \cdot \bf B=0$ requires $\bf B \cdot \nabla \alpha=0$. Note that $ \alpha=\bf B\cdot \nabla \times \bf B/{\bf B}^2$. The $\alpha$ parameter represents magnetic field twist. If $\alpha$ is too large the field becomes unstable. The dimensions of $\alpha \sim  1/L$ where L is a characteristic length scale of the field. The relative alpha is defined as the linear coefficient for the fitting of relative magnetic helicity and the magnetic free energy $E_f$ and magnetic flux $\Phi$, which is shown in Eq.~(\ref{eq:EFRMH}).
\begin{equation}
E_f \Phi ^2  = KH_R^2 \label{eq:EFRMH},
\end{equation}
where the physical dimension of the linear fitting coefficient $K$ is $L^{-1}$, just like the $\alpha$ in the force-free magnetic field ($\nabla\times\bf{B}=\alpha \bf{B})$. Hence, we name $K$ the relative alpha. Magnetic free energy $E_f$ is defined as the the difference between total magnetic energy in the volume and the magnetic energy of the potential field, which share the same normal magnetic field component at the lower boundary.
The relative magnetic helicity ($H_{\rm R}$) given by \citep{finn85}:
\begin{equation}
H_{\rm R} = \int {\left( {\mathbf A + \mathbf A_{\rm p} } \right)
\cdot \left( {\mathbf B - \mathbf P} \right)} dV,
\label{eq:RHvolume}
\end{equation}
which is gauge-invariant and we have choosen the potential magnetic field ($\bf{P}$) as the reference field.

\begin{figure*}[]
\centering
\includegraphics[angle=0,scale=0.4]{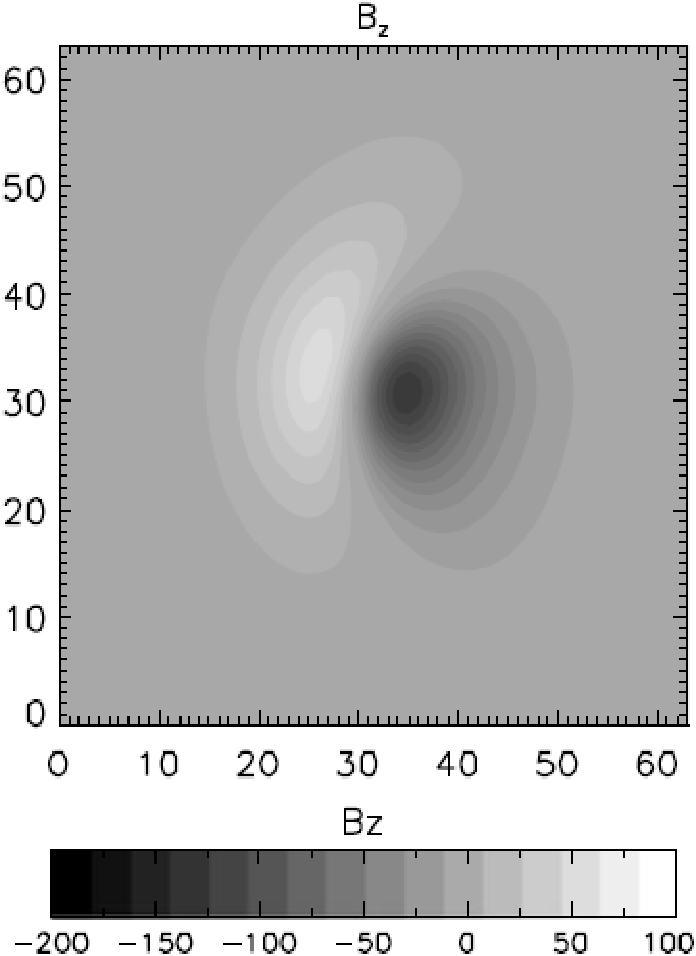}
\caption{Normal component of the magnetic field at the bottom boundary used for the constant alpha-force free field extrapolations.}
\label{fig:loub}
\end{figure*}

\begin{figure*}[]
\centering
\includegraphics[angle=0,scale=0.5]{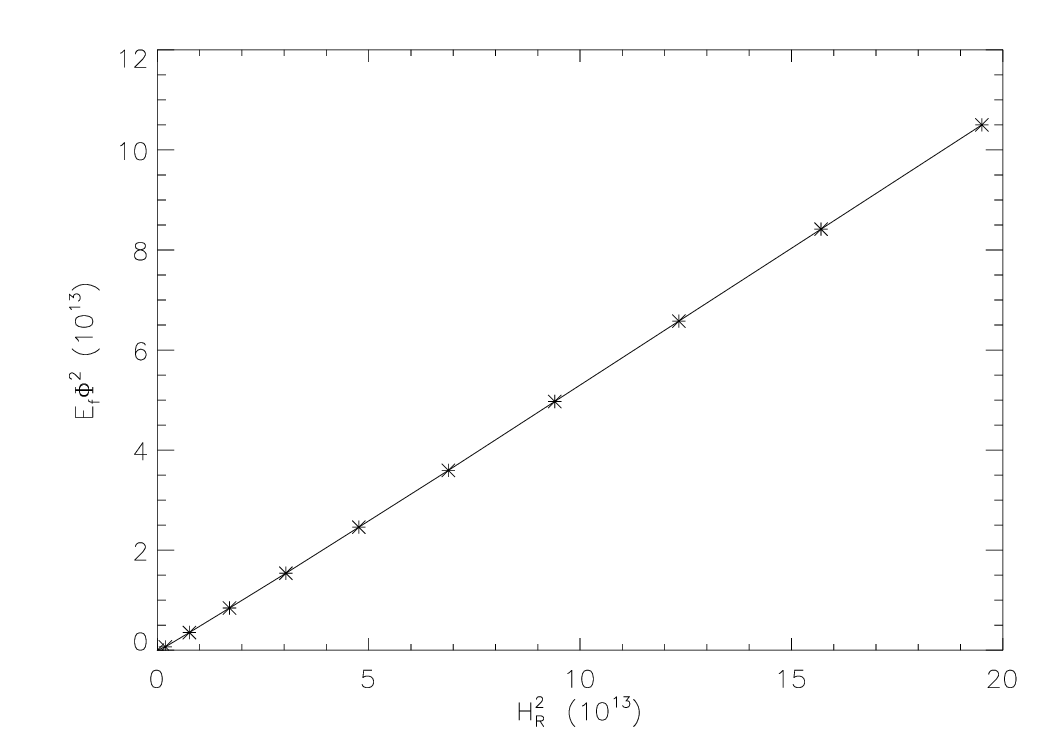}
\caption{Magnetic free energy versus relative magnetic helicity for the constant alpha-force free field ($\alpha\in[0,0.02]$) based on extrapolation method. The solid line represents a linear fit to the data.}
\label{fig:lou}
\end{figure*}

To calculate the relative alpha, we use the nonlinear force-free fields of \citet{Low90} as a sample. We choose the model labeled $P_{1,1}$ with $l = 0.3$ and $\Phi = \pi/4$ in the notation of their paper and the magnetic field is extrapolated in a grid of $64\times 64\times 64$. The methods of \citet{Ali81} and \citet{Gary89} is used to deduce the constant $\alpha$-force-free field based on the magnetic field at the bottom boundary (see Fig.~\ref{fig:loub}). The constant $\alpha$ is in the range of [0,0.02], with a step of 0.002. We calculate the relative magnetic helicity base on the method of \citet{Yang13,Yang18}, using the magnetic free energy and magnetic flux as described in Eq.~(\ref{eq:EFRMH}). Figure~\ref{fig:lou} shows the relation between the $E_f\Phi^2$ and ${H_R}^2$ for the given $\alpha$. It is clear that a good linear relation exists and the obtained relative alpha is $K=0.54$.

\section{Testing of simulations and observations}
\label{sec:test}

\subsection{Simulation Testing}

In this section, we used the output of two data-driven magnetohydrodynamics (MHD) models, LINMOD3D \citep{Santos08} and GOEMHD3 \citep{Skala15}, to test the proposed relation formula. The LINMOD3D model was used to investigate the plasma and magnetic field evolution in the solar atmosphere above active region NOAA~8210 in response to the energy influx from the lower boundary due to the horizontal plasma motions. The model starts with a potential field extrapolation obtained from the observed line-of-sight component of the photospheric magnetic field and plasma density and temperature close to the ones found in the solar atmosphere. The horizontal plasma motions inject magnetic helicity and free magnetic energy, which appears in the form of eletric currents. The currents are then dissipated by switching on an anomalous resistivity where the current carrier velocity is larger than the electron thermal velocity. The GOEMHD3 model is a massively parallel code for solving second-order accurate MHD equations. It is parallelized based on a hybrid MPI-OpenMP programming paradigm adopting a standard 2-D domain-decomposition approach. The weak Joule current dissipation and a finite viscosity are taken into account in the almost dissipationless solar corona. The model was used to investigate the long-time evolution of the relative magnetic helicity in the solar corona above active region NOAA~11249, both in ideal and non-ideal magnetohydrodynamics by the non-ideal magnetohydrodynamical phase triggered by the switching on of resistivity.

\begin{figure*}[]
\centering
\includegraphics[angle=0,scale=1.20]{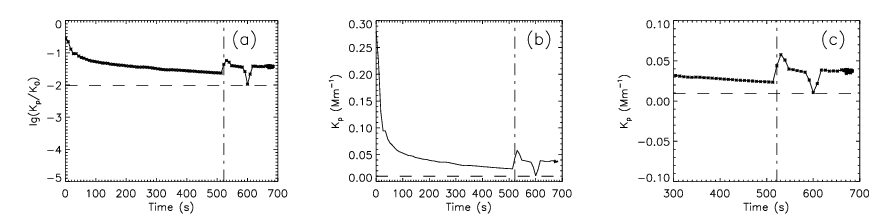}
\caption{(a) Temporal evolution of $K_p/K_0$ in logarithmic scale for NOAA~8210 ($K_0=1Mm^{-1}$).
(b) Temporal evolution of the parameter $K_p$ for NOAA~8210. (c) Close up near the time when magnetic reconnection initiated. The dash-dotted line represents the time when the fast reconnection started (t=522s). The dashed line represents the calculated coefficient $K_E$ based on Eq.~(\ref{eq:EFRMH}).}
\label{fig:8210}
\end{figure*}

\begin{figure*}[]
\centering
\includegraphics[angle=0,scale=1.20]{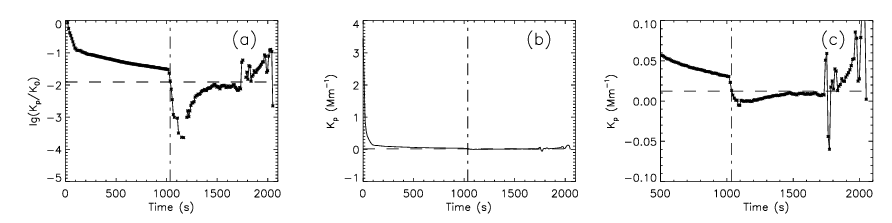}
\caption{(a) Temporal evolution of $K_p/K_0$ in logarithmic scale for NOAA~11429 ($K_0=1Mm^{-1}$).
(b) Temporal evolution of the parameter $K_p$ for NOAA~11429. (c) Close up near the time when magnetic reconnection initiated. The dash-dotted line represents the time when the magnetic flux eruption initiated (t=1035s). The dashed line represents the calculated coefficient $K_E$ based on Eq.~(\ref{eq:EFRMH}).}
\label{fig:11429}
\end{figure*}

The helicity evolution associated to active region NOAA~8210 was already studied in \citet{Yang13}, while the magnetic flux eruption of NOAA~11249 was investigated in \citet{Yang18}. Here we use the output of both simulations for evaluating the relation between relative magnetic helicity and free magnetic energy. We obtain the parameter $K_p(t)$ from:
\begin{equation}
K_p(t)  =\frac{{dE_f(t)\phi(t)^2}}{{dH^2_R(t)}}.
\label{eq:KPT}
\end{equation}
The evolution of $K_p$ for active regions NOAA~8210 and NOAA~11249 is shown in Fig.~\ref{fig:8210} and Fig.~\ref{fig:11429}, respectively. We also calculate a coefficient $K_E$ using Eq.~(\ref{eq:EFRMH}) by calculating the relative magnetic helicity and free magnetic energy based on the boundary condition of the initial magnetic field data, like described in Sec.~\ref{sec:EFHR}.

Figure~\ref{fig:8210} shows the evolution of $K_p$ for active region NOAA~8210. The horizontal dashed line represents the calculated coefficient $K_E$, based on Eq.~(\ref{eq:EFRMH}), while the vertical dash-dotted line represents the time when the fast reconnection started (t=522s). Figure~\ref{fig:8210}a shows the evolution of normalized $K_p/K_0$ in logarithmic scale, Fig.~\ref{fig:8210}b shows the temporal evolution of parameter $K_p$ and Fig.~\ref{fig:8210}c is a close up of parameter $K_p$ near the time when magnetic reconnection initiated. For active region NOAA
~8210, $K_E$ is equal 0.00953 $Mm^{-1}$. It is found that the $K_p$ evolve towards to the predicted value of $K_E$, as it was expected. At around time=600s, the $K_p$ values are very close to the value of $K_E$. In Fig.4c of \cite{Yang13}, the magnetic helicity change rate across the boundary also became constant after the time=600s, which means the driven force at right side of the momentum equation is also near zero, i.e. one extreme state.

Figure~\ref{fig:11429} presents the evolution of $K_p$ for active region NOAA~11429. The horizontal dashed line represents the calculated coefficient $K_E$ based on Eq.~(\ref{eq:EFRMH}). The vertical dash-dotted line represents the time when the fast reconnection started (t=1035s). Figure~\ref{fig:11429}a shows the $K_p$ evolution in logarithmic scale, Fig.~\ref{fig:11429}b shows the temporal evolution of $K_p$ and Fig.~\ref{fig:8210}(c) a close up near the time when magnetic reconnection initiated. For NOAA~11429, $K_E$ is equal 0.0123 $Mm^{-1}$ and, like for NOAA~8210, the $K_p$ also evolves toward to the predicted value of $K_E$. The $K_p$ is equal 0.0133 at time=1035s, which is very close to the predicted $K_E$ when the magnetic flux erupted.

\subsection{Observation Testing}

NOAA~11158 was investigated in many recent works for different purposes \citep[see,e.g.][]{Schrijver2011,Sun2012,Liu2013,Jing2012,Wiegelmann2012,Jiang2013}. This region is associated to 29 C-class, 1 M-class and 1 X-class solar flares during its passage through the solar disk. We use nonlinear force-free (NLFF) extrapolation method to obtain a 3D magnetic field associated to this active region and calculate the relative magnetic helicity evolution using the method of \citet{Yang13,Yang18}. In order to analyze the parameter $K_p$ while the solar flares happened in the studied active region, we define a weighted reference time $t_{ref}$ as follows:
\begin{equation}\label{eq:weight_time}
 {{t_{ref}} = \frac{{\sum {{t_i}SF{I_i}} }}{{SFI}},}
\end{equation}
where $t_i$ is the start time of considered solar flare event in the studied active region and $SFI_i$ is the corresponding solar flare index number of the solar flare. The SFI parameter is the total solar flare index of $SFI_i$ Eq~(\ref{eq:SFI}). This parameter is defined by Antalova (1996) to evaluate the released magnetic energy by weighting the SXR flares of classes B, C, M, and X as 0.1, 1, 10, and 100, respectively (in units of $10^{-6}W~m^{2}$ ):
\begin{equation}
  {SFI = \sum {(1000X + 100M + 10C + 1B + 0.1A)}.}
  \label{eq:SFI}
\end{equation}

\begin{figure*}[]
\centering
\includegraphics[angle=0,scale=1.2]{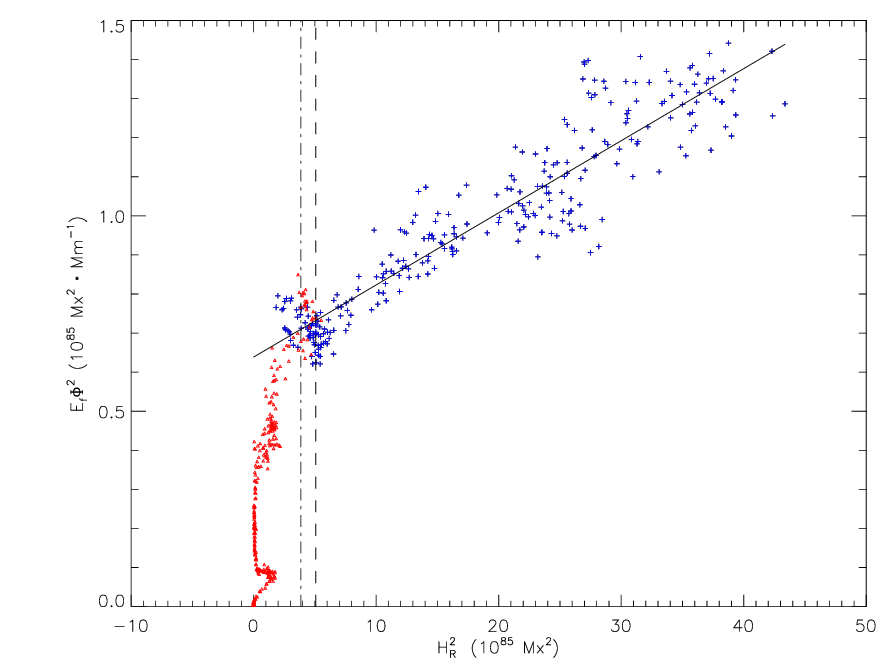}
\caption{Time profile of $K_p$ for NOAA11158. The plot points before the weighted time $t_{ref}$ at 07:44UT on Feb. 15, 2011 are labeled by red asterisk sign. The plot points after $t_{ref}$ are labeled by blue cross sign. The vertical dash-dotted line represents the squared relative magnetic helicity ($H_R^2$) at  ${t_{ref}}$. The vertical dash-dotted line represents the squared relative magnetic helicity when the X2.2 solar flare happened at 01:44UT on Feb. 15, 2011. The solid line is a linear fitting of $H_R^2$ and $E_f\Phi^2$ for the blue points after $t_{ref}$.}
\label{fig:11158}
\end{figure*}

For NOAA~11158, the weighted time $t_{ref}$ is at 07:44~UT on February 15 2011. The X2.2 Class solar flare happened at 01:44~UT on the same day, which is a little earlier than the weighted time. This happens because there is a M-class flare (M6.6) on February 13 which moves the barycenter of time weighted by the solar flare index further in time. Fig.~(\ref{fig:11158}) depicts the time evolution of $K_p$ for NOAA~11158. The plot points before (after) the weighted time $t_{ref}$ at 07:44~UT on February 15 2011 are labeled by red (blue) asterisk (cross) signs. The vertical dash-dotted line represents the squared relative magnetic helicity ($H_R^2$) at ${t_{ref}}$. The vertical dash-dotted line represents the squared relative magnetic helicity when the X2.2 solar flare happened at 01:44~UT on February 15 2011. The solid line is a linear fitting of $H_R^2$ and $E_f\Phi^2$ for the blue points after $t_{ref}$ and the linear fitting slope ($K_p$) is 0.0185 $Mm^{-1}$. We also calculate the coefficient $K_E$ using Eq.~(\ref{eq:EFRMH}) by calculating the relative magnetic helicity and free magnetic energy based on the photospheric magnetic field of NOAA~11158. Note that this active reigon is an emerging active region. We choose the photospheric magnetic field of NOAA~11158 at 01:36UT on February 15 2011 just before the X2.2 solar flare occurred. The calculated $K_E$ is 0.0196 $Mm^{-1}$, which is very close to the linear fitting value (difference less than 7\%). This consistence implies the calculation of $K_E$ based on the solar photospheric magnetic field measurement could play an important role to predict the existence of MHD extreme state just before the strong release process of magnetic free energy.

\section{Summary}
\label{sec:summary}

In this work, a coefficient called relative alpha between the relative magnetic helicity and free magnetic energy in the MHD extreme state is proposed with the open magnetic field boundary. We present the calculation process of relative alpha using the theoretical nonlinear force-free model and test it in the output of an MHD mode for  NOAA~8210 and NOAA~11429, and also 3D magnetic field data of NOAA
~11158 obtained through NLFF extrapolation approach based on the HMI/SDO data. It was found that the defined relative alpha in both cases evolve toward to the calculated coefficient $K_E$. We propose the use of the evolution of $K_p(t)$ to evaluate whether the solar eruptions will happen. This will be of great help to predict the solar activity like the solar flares and CMEs for space weather forecast.

According to Eq.~(\ref{eq:EFRMH}), ${{dH_R}}/{{dE_f}}$ is proportional to $E_f^{-0.5}$ if we fixed normal part of magnetic field in the MHD extreme state.  As the free magnetic field energy is injected into this considered volume, the change rate ${{dH_R}}/{{dE_f}}$ tends to be zero and the accumulated magnetic  helicity will tend to be constant just before solar eruption. \citet{Park08} investigated the variation of magnetic helicity for the 11 X-class flares which occurred in seven ARs. They found that each of these major flares was preceded by a significant helicity accumulation at a nearly constant rate and then becomes nearly constant before the flares. These observed phenomena could be explained naturally by our theory. This also explains the findings of  \citet{ZhangM06,ZhangM08} that conjectured there is an upper-limit of the total magnetic helicity that a force-free magnetic field can contain before it erupts.
.

\begin{acknowledgements}
\small
 This research is supported by the National Key $R\&D$ Program of China No. 2022YFF0503800 and 2021YFA1600500; National Natural Science Foundation of China (grants No. 12250005, 12073040, 11973056, 12003051, 11573037, 12073041, 11427901, and 11611530679); by the Strategic Priority Research Program of the Chinese Academy of Sciences (grants No. XDB0560000, XDA15052200, XDB09040200, XDA15010700, and XDA15320102); Natural Science Foundation of Beijing (Grant No. Z180007); by the Youth Innovation Promotion Association of CAS (2019059).The authors would  also like to thank the Supercomputing Center of the Chinese Academy of Sciences (SCCAS).; by the Max-Planck Society Interinstitutional Research Initiative Turbulent transport and ion heating, reconnection, and electron acceleration in solar and fusion plasmas of Project No. MIF-IF-A-AERO8047; by Max-Planck-Princeton Center for Plasma Physics PS AERO 8003; by ISSI International Team on Magnetic Helicity estimations in models and observations of the solar magnetic field. The authors would  also like to thank the Supercomputing Center of the Chinese Academy of Sciences (SCCAS) and the  Max-Planck Computing and Data Facility (MPCDF) for the allocation of computing time.
\end{acknowledgements}

\end{document}